# Reduction of CRB in Arbitrary Pre-designed Arrays Using Alter an Element Position

Mohammadreza darabi

Nahavand cement company, *Tehran, Iran*
*E-mail:* mr.darabi@nahavandcement.com

*Abstract*— **Simultaneous estimation of range and angle of close emitters usually requires a multidimensional search. This paper offers an algorithm to improve the position of an element of any array designed on the basis of some certain or random rules. In the proposed method one element moves on its original direction, i.e., keeping the vertical distance to each source, to reach the constellation with less CRB. The performance of this method has been demonstrated through simulation and a comparison of the CRB with receptive signals covariance matrix determinant has been made before and after the use of this method.**

## I. INTRODUCTION

Direction of arrival (DOA) estimation usually is found under the assumption that signal sources are in the far field of the array, and hence the wavefront is plannar across the array aperture. Suppose that far field range is defined as $R_0$, so that the range at which the largest departure of the wavefront from a plane wave, across the array, is $l\lambda$, where $\lambda$ is the wavelength. Then it is straightforward to show that $R_0 \approx D^2/8l$, where $D$ is the array aperture measured in wavelengths [1]. For arrays with small aperture, $R_o$ is rather small and the far field assumption holds very well. However, for arrays with large aperture, e.g., those used in sonar systems, sources are usually located in the near field [2].

Bearing estimation for near-field sources requires simultaneous estimation of the bearing and range, that is because the curvature of wavefront cannot be ignored. This estimation usually requires a multidimensional search.

Previous works on this estimation can be seen in [1-6]. Starer and Nehorai [5] developed an algorithm based on path-following. This algorithm is limited to uniform linear arrays and to sources that are located in the Fresnel region. This region is taken to be between near-field case (having spherical wavefronts) and far-field case (where wavefronts can be estimated in plane form). Collins et al. [6] offer an analytic simulated algorithm to solve the problem of estimating the range and bearing. Their algorithm is also limited to uniform linear arrays.

The works in [7-10] have also focused on transforming the problem of single dimensional search into polynomial rooting in different cases. The aim of this paper is not to offer a constellation, but to change the position of an element of any array designed on the basis of some certain or random rules to reach the constellation with less Cramer-Rao bound (CRB). In section II, received signal is modeled as a deterministic signal in AWGN. Section III, briefly states previously obtained results on CRB. Section IV, describes changing the position of an element to obtain less CRB. Section V evaluates theory results using computer simulation. In section VI, optimization procedure has been summarized and results have been presented.

## II. SYSTEM MODEL

Let us take $N$ sources of emitters that are observed by arbitrary array of $M$ sensors. The signal at the output of the $m$th sensor can be described by:

$$x_m(t) = \sum_{n=1}^{N} s_n(t - \tau_{mn}) + v_m(t); \qquad -T/2 \le t \le T/2, \quad m = 1, 2, \ldots, M \tag{1}$$

where $\{s_n(t)\}_{n=1}^{N}$ are the radiated signals, $\{v_m(t)\}_{m=1}^{M}$ are waveforms of additive noise processes and $T$ is the observation interval. The parameter $\tau_{mn}$ is the delay associated with the signal propagation time from the $n$th source to the $m$th sensor. These parameters are focused upon since they contain information about the position of source to arrays.

Applying appropriate Fourier transform on (1), results:

$$X_m(j) = \sum_{n=1}^{N} e^{-j\omega_{0n}\tau_{mn}} S_n(j) + V_m(j); \qquad m = 1, 2, \ldots, M \tag{2}$$

where $S_n(j)$ and $V_m(j)$ are Fourier transform of $s_n(t)$ and $v_m(t)$, respectively, and $\omega_{0n}$ denotes center frequency of $n$th source radiated signal. The index $j$ represents the snapshot number. Using vector notations, formula (2) can be restated as follows:

$$\underline{X}(j) = \mathbf{A}\underline{S}(j) + \underline{V}(j) \tag{3}$$

where

$$
\begin{aligned}
\underline{X}(j) &= \left[X_1(j), X_2(j), \ldots, X_M(j)\right]^T \\
\underline{S}(j) &= \left[S_1(j), S_2(j), \ldots, S_N(j)\right]^T \\
\underline{V}(j) &= \left[V_1(j), V_2(j), \ldots, V_M(j)\right]^T \\
\mathbf{A}_{mn} &= \left[e^{-j\omega_0\tau_{mn}}\right]_{M \times N} \quad; m = 1, 2, \ldots, M, n = 1, 2, \ldots, N
\end{aligned}
\tag{4}
$$

To simplicity, we assume that sensors and sources are located on the same common plane, so [10]:

$$\tau_{mn} = \frac{1}{c} r_n \left[1 + \left(\frac{\rho_m}{r_n}\right)^2 - 2\frac{\rho_m}{r_n}\cos(\theta_n - \phi_m)\right]^{1/2} \tag{5}$$

where $c$ is the propagation velocity, $r_n$ and $\theta_n$ are the range and bearing of the $n$th source, and $p_m$, $\phi_m$ are the polar coordinates of the $m$th sensor. The problem is the estimation of $\{\underline{r}_n, \underline{\theta}_n\}_{n=1}^{N}$ using data $\{\underline{X}(j)\}_{j=1}^{N_s}$, where $N_s$ is the number of snapshots.

Before we obtain CRB, defining the following covariance matrices is necessary:

$$\mathbf{R}_s = E\left\{\underline{S}(j)\underline{S}^H(j)\right\} \tag{6}$$

$$\mathbf{R}_n = E\left\{\underline{V}(j)\underline{V}^H(j)\right\} = \eta\mathbf{I} \tag{7}$$

$$\mathbf{R}_x = E\left\{\underline{X}(j)\underline{X}^H(j)\right\} = \mathbf{A}\mathbf{R}_s\mathbf{A}^H + \mathbf{R}_n \tag{8}$$

We also apply the following sample covariance matrix as an estimation of $\mathbf{R}_x$ [2]:

$$\hat{\mathbf{R}}_x = \frac{1}{N_s}\sum_{j=1}^{N_s}\underline{X}(j)\underline{X}^H(j) \tag{9}$$

## III.   CRAMER RAO BOUND

In this section, we briefly explain CRB for the estimation of the DOA of the $N$ observed sources with $M$ elements. An array with $M$ elements can at most separate $M-1$ sources. Therefore, $N < M$ is needed.

CRB gives a lower bound on the covariance matrix of any unbiased estimator. Let's assume that we have $N_s$ independent samples of zero mean Gaussian process $x$ which statistically depend on an arbitrary parameter vector $\underline{P}$, and then the Fisher Information Matrix (FIM) is as follows [11]:

$$\mathbf{F}_{mn} = N_s.tr\left\{\mathbf{R}_x^{-1}\frac{\partial\mathbf{R}_x}{\partial p_m}\mathbf{R}_x^{-1}\frac{\partial\mathbf{R}_x}{\partial p_n}\right\} \tag{10}$$

CRB is equal to the main diagonal elements of $\mathbf{F}$ inverse, [11].

The following relation can be obtained from (8):

$$\mathbf{R}_x = \mathbf{A}\mathbf{R}_s\mathbf{A}^H + \eta\mathbf{I} \tag{11}$$

We assume that the signals and the Gaussian noise are uncorrelated. However, signals might be correlated or even coherent.

We define our parameter vector as follows:

$$\underline{P} = \left[\underline{\theta}^T, \underline{r}^T, \underline{\mu}^T, \nu\right]^T \tag{12}$$

where $\underline{\theta}$ is the DOA vector of $N$ signals, $\underline{r}$ is the corresponding vectors of $N$ source ranges, $\underline{\mu}$ is a parameter vector that specifies the $\mathbf{R}_s$ entries, and $\nu$ denotes the noise variance. FIM can be partitioned into two blocks, each of which is linked to one or two parametric vector in $\underline{P}$. It is shown that those blocks are as follows [2]:

$$\mathbf{F}_{\underline{\alpha}\underline{\beta}} = 2\operatorname{Re}\left\{ \left(\mathbf{R}_s \mathbf{A}^H \mathbf{R}_x^{-1} \mathbf{A} \mathbf{R}_s\right) \times \left(\dot{\mathbf{A}}_{\underline{\beta}}^H \mathbf{R}_x^{-1} \dot{\mathbf{A}}_{\underline{\alpha}}^H\right)^T + \left(\mathbf{R}_s \mathbf{A}^H \mathbf{R}_x^{-1} \dot{\mathbf{A}}_{\underline{\beta}}\right) \times \left(\mathbf{R}_s \mathbf{A}^H \mathbf{R}_x^{-1} \dot{\mathbf{A}}_{\underline{\alpha}}\right)^T \right\} \tag{13}$$

$$\mathbf{F}_{\underline{\alpha}\underline{\mu}} = \mathbf{Q}_4 = \left\{ \left(\mathbf{A}^H \mathbf{R}_x^{-1} \mathbf{A} \mathbf{R}_s\right)^T \otimes \left(\dot{\mathbf{A}}_{\underline{\alpha}}^H \mathbf{R}_x^{-1} \mathbf{A}\right) \qquad = \left(\mathbf{A}^H \mathbf{R}_x^{-1} \dot{\mathbf{A}}_{\underline{\alpha}}\right)^T \times \left(\mathbf{R}_s \mathbf{A}^H \mathbf{R}_x^{-1} \mathbf{A}\right) \right\} \mathbf{Q}_t^H \tag{14}$$

$$\mathbf{F}_{\underline{\alpha}\nu} = 2\operatorname{Re}\left\{ diag\left(\mathbf{R}_s \mathbf{A}^H \mathbf{R}_x^{-2} \dot{\mathbf{A}}_{\underline{\alpha}}\right) \right\} \tag{15}$$

$$\mathbf{F}_{\underline{\mu}\underline{\mu}} = \mathbf{Q}_t \left[ \left(\mathbf{A}^H \mathbf{R}_x^{-1} \mathbf{A}\right)^* \otimes \left(\mathbf{A}^H \mathbf{R}_x^{-1} \mathbf{A}\right) \right] \mathbf{Q}_t^H \tag{16}$$

$$\mathbf{F}_{\underline{\mu}\nu} = \mathbf{Q}_t \left[ \left(\mathbf{R}_x^{-1} \mathbf{A}\right)^T \otimes \left(\mathbf{R}_x^{-1} \mathbf{A}\right)^H \right] \underline{I}(:) \tag{17}$$

$$F_{\nu\nu} = tr\left\{ \mathbf{R}_x^{-2} \right\} \tag{18}$$

where $\underline{\alpha}$ and $\underline{\beta}$ can be either one of the $\underline{\theta}$ and $\underline{r}$ parameter vectors. We used $\times$ here to denote Hadamard product of two. Also, $\otimes$ shows Kronecker product [12-13]. $\underline{I}(:)$ denotes a vector consisting of a concatenation of the columns of the matrix and $diag\{B\}$ is a column vector of containing the diagonal elements of the matrix $B$. The operators $(\ )^T, (\ )^*$ and $(\ )^H$ represent transposition, conjugation, and conjugate transposition, respectively. The appendix describes details of the above equations.

Formulas (13) to (18) provide a set of closed form of formulas to compute two dimensional CRB (2-D CRB).

## IV. FINDING THE OPTIMAL CONSTELLATION TO MINIMIZE CRB

This section deals with an element position. In fact, this element is which element with the strongest receptive signal. The aim of this section is not to offer a constellation, but to modify the position of an element in a pre-designed arbitrary array, in a way that one element moves in the same direction, i.e., keeps a vertical distance to each source, to reach a constellation with the less CRB.

We know that [8]:

$$\left| \det\left([\mathbf{A}]_{M \times M}\right) \right|_{abs} \leq \left\{ \max \left|a_{ij}\right|_{abs} \right\}^M M^{M/2} \; ; \qquad i, j = 1, 2, \ldots, M \tag{19}$$

So using the above formula, without loss of generality, we can take the largest entry of $\mathbf{R}_x$ as the same $k$th element. In other words, it is possible to assign the number of $k$ to an element with the strongest receptive signal. We have:

$$\left|\det \mathbf{R}_x\right|_{abs} \le \left\{E\left\{\underline{X}_k(j)\underline{X}_k^*(j)\right\}\right\}^M M^{M/2} = \left(\left|\underline{X}_k(j)\right|^2\right)^M M^{M/2} \tag{20}$$

To reduce determinant of $\mathbf{R}_x$, $\left|X_k(j)\right|^2$ should be reduced. Therefore:

$$\left|X_k(j)\right|^2 = \left|\sum_{n=1}^{N} S_n(j)(\cos \omega_{0n}\tau_{kn} + j\sin \omega_{0n}\tau_{kn})\right|^2 = \left|\sum_{n=1}^{N}(S_n(j)\cos \omega_{0n}\tau_{kn}) + j\sum_{n=1}^{N}(S_n(j)\sin \omega_{0n}\tau_{kn})\right|^2$$

$$= \left(\sum_{n=1}^{N}(S_n(j)\cos \omega_{0n}\tau_{kn})\right)^2 + \left(\sum_{n=1}^{N}(S_n(j)\cos \omega_{0n}\tau_{kn})\right)^2 \tag{21}$$

Since $S_n(j); \; n = 1, 2, \ldots, N$, depend on sources specification, then they don't have significant role on determinant minimizing. If $S_l(j)$ be the largest $S_n(j)$, then we have:

$$\left|X_k(j)\right|^2 \le \left(S_l(j)\sum_{n=1}^{N}\cos \omega_{0n}\tau_{kn}\right)^2 + \left(S_l(j)\sum_{n=1}^{N}\sin \omega_{0n}\tau_{kn}\right)^2 = S_l^2(j)\left\{G + F\right\} \tag{22}$$

where

$$G = \left(\sum_{n=1}^{N}\cos \omega_{0n}\tau_{kn}\right)^2 \quad ; n = 1, 2, \ldots, N \tag{23}$$

$$F = \left(\sum_{n=1}^{N}\cos \omega_{0n}\tau_{kn}\right)^2 \quad ; n = 1, 2, \ldots, N \tag{24}$$

Now something should be done to reduce the $\left\{G + F\right\}$. Fig. 1 shows an array with $M$ elements and $N$ sources so that $H_{mn}$ and $\varphi_{mn}$ denote vertical distance and angle of arrival between $m$th element and $n$th source. According to this figure we have:

$$\sin \varphi_{kn} = \frac{H_{kn}}{C\tau_{kn}} \quad ; n = 1, 2, \ldots, N \tag{25}$$

$$\tau_{kn} = \frac{H_{kn}}{C\sin \varphi_{kn}} \quad ; n = 1, 2, \ldots, N \tag{26}$$

where $C$ is transmit velocity of signals in an environment. Then:

$$G + F = \left\{\left(\sum_{n=1}^{N}\cos T_n\right)^2 + \left(\sum_{n=1}^{N}\sin T_n\right)^2\right\} \tag{27}$$

where

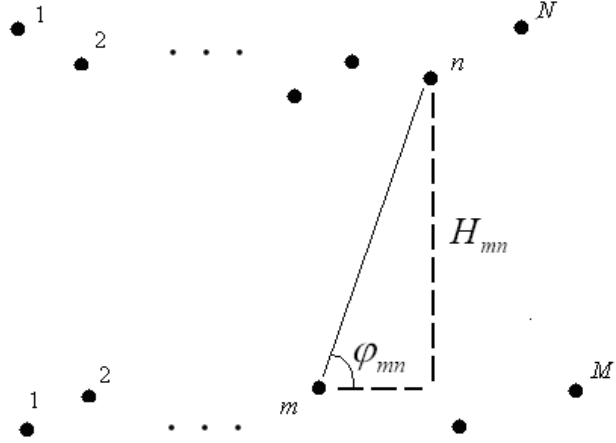

Fig. 1. Arbitrary constellation of sensors and sources

$$T_n = \frac{g}{\sin \varphi_{kn}} \quad ; n = 1, 2, \ldots, N \tag{28}$$

$$g = \frac{\omega_{0n} H_{kn}}{C} \quad ; n = 1, 2, \ldots, N \tag{29}$$

Thus

$$
\begin{aligned}
G + F &= \left(\cos T_1 + \cos T_2 + \cdots + \cos T_N\right)^2 + \left(\sin T_1 + \sin T_2 + \cdots + \sin T_N\right)^2 \\
&= \cos^2 T_1 + \cos^2 T_2 + \cdots + \cos^2 T_N \\
&\quad + \sin^2 T_1 + \sin^2 T_2 + \cdots + \sin^2 T_N \\
&\quad + 2\cos T_1 \cos T_2 + \cdots + 2\cos T_{N-1} \cos T_N \\
&\quad + 2\sin T_1 \sin T_2 + \cdots + 2\sin T_{N-1} \sin T_N \\
&= N + 2\left(\cos T_1 \cos T_2 + \cdots + \cos T_{N-1} \cos T_N + \sin T_1 \sin T_2 + \cdots + \sin T_{N-1} \sin T_N\right)
\end{aligned}
\tag{30}
$$

The above equation should be reduced.

We know that the minimum of a phrase contains the summation of cross-product of sinusoidal phrase and cosine phrase for the even number of angles occurs when the angles are equally $0$ and $\pi/2$, and for odd number of angles occurs when the angles are alternatively $0$ and $\pi/2$.

What the above paragraph means is that $T_n$; $n = 1, 2, \ldots, N$, should be equally zero and $\pi/2$. In other words, the number of $\lfloor N/2 \rfloor$ (where $\lfloor . \rfloor$ denote the small nearest integer) of $T_i$ should equal zero and the same number should be $\pi/2$. When $N$ is odd, then the number of zeroes or $\pi/2$ is one unit bigger and the increase of each of them will have identical effects in final results. So:

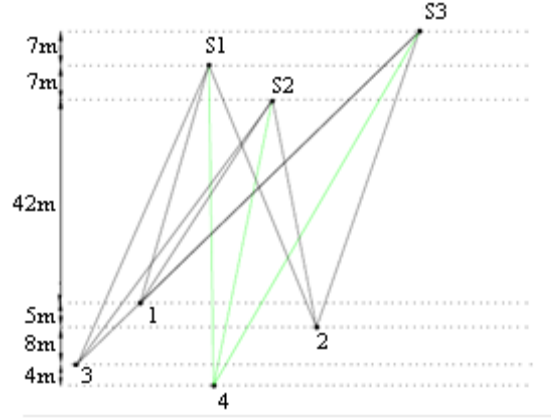

Fig. 2. Arbitrary designed array such that $\mathbf{R}_x$ determinant is equal to $1.8112 \times 10^{-42}$

$$T_n = 0 \ , \quad T_{n+1} = \frac{\pi}{2} \ ; \qquad n = 1, 2, \ldots, N \tag{31}$$

$$\frac{\omega_{0n} H_{kn}}{C \sin \varphi_{kn}} = 0 \quad , \quad \frac{\omega_{0(n+1)} H_{k(n+1)}}{C \sin \varphi_{k(n+1)}} = \frac{\pi}{2} \tag{32}$$

Using a simple set of algebraic phrases and knowing that $\omega_{0n} = 2\pi f_{0n}$, we will obtain:

$$\varphi_{kn} = arc \sin \frac{2m f_{0n} H_{kn}}{C} \tag{33}$$

$$\varphi_{k(n+1)} = arc \sin \frac{4 f_{0(n+1)} H_{k(n+1)}}{C} \tag{34}$$

where $m$ represents a number which, whenever it gets bigger, the estimation and determination of the position of arrays gets more accurate. In fact, since the $T_i$ ; $i = 1, 2, \ldots, N$, cannot be zero, we can take it as a small number like $\pi/m$.

## V. NUMERICAL RESULTS

### A. Simulation based on three sources

In this section, we evaluate theory results using computer simulation. The simulation is based on 3 sources and 4 elements. If the index $j$ represents the constant snapshot number, we consider baseband transmit signals as $S_1(j) = 2 + i2$, $S_2(j) = 1 + i3$, $S_3(j) = 5 + i3$, where $i = \sqrt{(-1)}$, and also frequency and propagation velocity are as $f_{01} = 1.1787 \times 10^6 Hz$, $f_{02} = 10^4 Hz$, $f_{03} = 9.9298 \times 10^5 Hz$, $C = 3 \times 10^8 \, m/s$, respectively. Fig. 2 corresponds to the array elements DOAs and vertical ranges according to table 1 and table 2 values. So we compute the $\mathbf{R}_x$ determinant, $CRB_\theta$ and $CRB_r$. These values are equal to $1.8112 \times 10^{-42}$, $1.0308 \times 10^{-29}$, and $8.8505 \times 10^{-26}$, respectively. Then, to evaluate the effect of change an element location, we compute the power of receptive signals in each element and observed that third element have the strongest receptive signal. Then we change array constellation in basis of section **IV**, so that new values of angels have been obtained as $\varphi_{31} = 103°$, $\varphi_{32} = 90°$, $\varphi_{33} = 66°$, Fig. 3. Then we compute the $\mathbf{R}_x$ determinant, $CRB_\theta$ and $CRB_r$ for proposed constellation and observed that these values are equal to $3.4102 \times 10^{-44}$, $6.0716 \times 10^{-31}$, and $6.1691 \times 10^{-27}$, respectively.

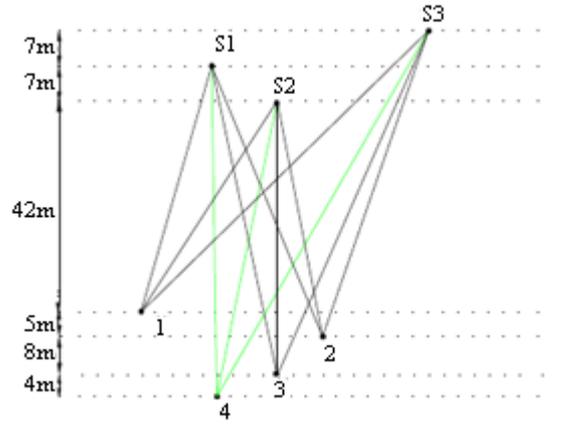

Fig. 3. Proposed constellation such that $\mathbf{R}_x$ determinant is equal to $3.4102 \times 10^{-44}$

Table 1. Angles between elements and source versus degree

| Sources<br>Elements | 1 | 2 | 3 |
|---|---|---|---|
| 1 | 74 | 56 | 45 |
| 2 | 111 | 100 | 70 |
| 3 | 66 | 53 | 44 |
| 4 | 90 | 77 | 59 |

Table 2. Vertical distances between elements and sources versus meter

| Sources<br>Elements | 1 | 2 | 3 |
|---|---|---|---|
| 1 | 49 | 42 | 56 |
| 2 | 54 | 47 | 61 |
| 3 | 62 | 55 | 69 |
| 4 | 66 | 59 | 73 |

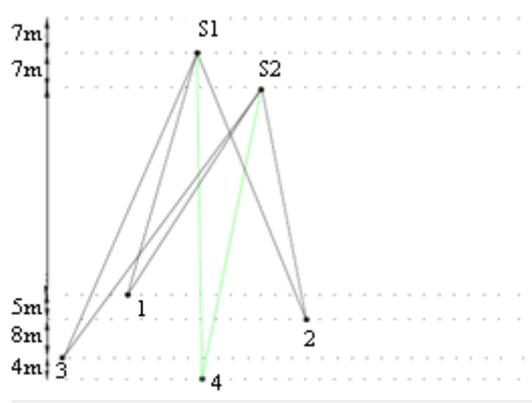

Fig. 4. Arbitrary designed array using two sensors

Table 3. Angles between elements and sources versus degree

| Sources / Elements | 1 | 2 |
|---|---|---|
| 1 | 74 | 56 |
| 2 | 111 | 100 |
| 3 | 66 | 53 |
| 4 | 90 | 77 |

Table 4. Vertical distances between elements and sources versus meter

| Sources / Elements | 1 | 2 |
|---|---|---|
| 1 | 49 | 42 |
| 2 | 54 | 47 |
| 3 | 62 | 55 |
| 4 | 66 | 59 |

*B. Simulation based on two sources*

Simulations in this section deal with two sources, and results have been evaluated based on 2 sources and 4 elements. We consider transmit signals as $S_1(j) = 2+i2$, $S_2(j) = 1+i3$, and also frequency and propagation velocity are as $f_{01} = 1.5 \times 10^6 Hz$, $f_{02} = 9 \times 10^5 Hz$, $C = 3 \times 10^8 \, m/s$, respectively.

Fig. 4 corresponds to the array elements DOAs and vertical ranges according to table 3 and table 4 values. So we compute the $\mathbf{R}_s$ determinant, $CRB_\theta$ and $CRB_r$. These values are equal to $2.3737 \times 10^{-42}$, $5.0207 \times 10^{-30}$, and $4.3154 \times 10^{-30}$, respectively.

TABLE 5. $\boldsymbol{R}_x$ DETERMINANT VERSUS FREQUENCY FOR AN ARRAY WITH FOUR ELEMENTS (M=4)

| Points (Hz) | Primary | | An Element Changing | |
|---|---|---|---|---|
| | 2 Sources | 3 Sources | 2 Sources | 3 Sources |
| 1000000 | $8.25 \times 10^{-42}$ | $5.31 \times 10^{-42}$ | $3.41 \times 10^{-42}$ | $3.61 \times 10^{-42}$ |
| 2000000 | $2.05 \times 10^{-42}$ | $5.83 \times 10^{-43}$ | $1.81 \times 10^{-42}$ | $4.39 \times 10^{-42}$ |
| 3000000 | $1.57 \times 10^{-43}$ | $2.82 \times 10^{-42}$ | $1.10 \times 10^{-43}$ | $1.95 \times 10^{-432}$ |
| 4000000 | $1.07 \times 10^{-43}$ | $3.56 \times 10^{-42}$ | $3.64 \times 10^{-43}$ | $6.16 \times 10^{-43}$ |
| 5000000 | $1.31 \times 10^{-43}$ | $4.12 \times 10^{-42}$ | $1.63 \times 10^{-43}$ | $4.00 \times 10^{-43}$ |
| 6000000 | $4.12 \times 10^{-43}$ | $3.12 \times 10^{-42}$ | $3.22 \times 10^{-44}$ | $8.50 \times 10^{-43}$ |
| 7000000 | $1.78 \times 10^{-43}$ | $1.30 \times 10^{-42}$ | $6.97 \times 10^{-44}$ | $1.99 \times 10^{-43}$ |
| 8000000 | $3.95 \times 10^{-43}$ | $1.15 \times 10^{-43}$ | $3.25 \times 10^{-45}$ | $2.43 \times 10^{-44}$ |
| 9000000 | $8.67 \times 10^{-44}$ | $7.00 \times 10^{-42}$ | $7.66 \times 10^{-45}$ | $1.45 \times 10^{-44}$ |
| 10000000 | $4.52 \times 10^{-44}$ | $1.58 \times 10^{-42}$ | $1.52 \times 10^{-45}$ | $8.03 \times 10^{-44}$ |

TABLE 6. $\boldsymbol{R}_x$ DETERMINANT VERSUS PROPAGATION VELOCITY FOR AN ARRAY WITH FOUR ELEMENTS (M=4)

| Points (m/s) | Primary | | An Element Changing | |
|---|---|---|---|---|
| | 2 Sources | 3 Sources | 2 Sources | 3 Sources |
| 1000000 | $5.97 \times 10^{-43}$ | $2.55 \times 10^{-41}$ | $1.94 \times 10^{-43}$ | $2.89 \times 10^{-43}$ |
| 2000000 | $1.10 \times 10^{-43}$ | $1.97 \times 10^{-41}$ | $6.71 \times 10^{-43}$ | $4.15 \times 10^{-43}$ |
| 3000000 | $1.81 \times 10^{-43}$ | $6.95 \times 10^{-41}$ | $3.90 \times 10^{-43}$ | $1.84 \times 10^{-43}$ |
| 4000000 | $3.77 \times 10^{-43}$ | $7.70 \times 10^{-42}$ | $3.19 \times 10^{-43}$ | $8.23 \times 10^{-43}$ |
| 5000000 | $3.57 \times 10^{-43}$ | $3.15 \times 10^{-42}$ | $3.31 \times 10^{-43}$ | $4.41 \times 10^{-42}$ |
| 6000000 | $7.74 \times 10^{-43}$ | $1.42 \times 10^{-42}$ | $1.03 \times 10^{-42}$ | $2.02 \times 10^{-42}$ |
| 7000000 | $1.18 \times 10^{-42}$ | $5.57 \times 10^{-42}$ | $6.65 \times 10^{-42}$ | $1.22 \times 10^{-42}$ |
| 8000000 | $1.39 \times 10^{-42}$ | $5.94 \times 10^{-43}$ | $6.04 \times 10^{-42}$ | $3.09 \times 10^{-42}$ |
| 9000000 | $2.42 \times 10^{-42}$ | $1.24 \times 10^{-43}$ | $2.16 \times 10^{-42}$ | $5.31 \times 10^{-42}$ |
| 10000000 | $8.73 \times 10^{-42}$ | $3.23 \times 10^{-43}$ | $5.58 \times 10^{-42}$ | $1.53 \times 10^{-42}$ |

Then to evaluate the effect of change an element location, we compute the power of receptive signals in each element and observed that third element have the strongest receptive signal. So we change array constellation in basis of section **IV**, so that new values of angels have been obtained as $\varphi_{31} = 90^\circ$, $\varphi_{32} = 41.3^\circ$. Then we compute the $\mathbf{R}_x$ determinant, $CRB_\theta$ and $CRB_r$ for proposed constellation and observed that these values are equal to $4.1649 \times 10^{-43}$, $1.1578 \times 10^{-30}$, and $2.1629 \times 10^{-30}$, respectively.

Tables 5 to 10 demonstrate the performance of above algorithm in comparison with primary constellation with altering the propagation velocity, and first source frequency, so that tables 5 and 6 show $\mathbf{R}_x$ determinant versus frequency and propagation velocity, respectively. Tables 7 and 8 demonstrate $CRB_\theta$ versus frequency and propagation velocity, and tables 9 and 10 show $CRB_r$ versus frequency and propagation velocity, respectively.

TABLE 7. $CRB_\theta$ VERSUS FREQUENCY FOR AN ARRAY WITH FOUR ELEMENTS (M=4)

| Points | Primary | | An Element Changing | |
|---|---|---|---|---|
| (Hz) | 2 Sources | 3 Sources | 2 Sources | 3 Sources |
| 1000000 | $2.27 \times 10^{-28}$ | $3.09 \times 10^{-30}$ | $3.50 \times 10^{-29}$ | $2.12 \times 10^{-30}$ |
| 2000000 | $2.77 \times 10^{-30}$ | $1.12 \times 10^{-30}$ | $1.01 \times 10^{-30}$ | $7.84 \times 10^{-30}$ |
| 3000000 | $5.32 \times 10^{-30}$ | $2.49 \times 10^{-30}$ | $2.51 \times 10^{-30}$ | $1.44 \times 10^{-30}$ |
| 4000000 | $1.47 \times 10^{-30}$ | $4.98 \times 10^{-31}$ | $2.23 \times 10^{-30}$ | $2.20 \times 10^{-31}$ |
| 5000000 | $4.44 \times 10^{-30}$ | $2.70 \times 10^{-31}$ | $4.23 \times 10^{-31}$ | $1.41 \times 10^{-31}$ |
| 6000000 | $1.06 \times 10^{-30}$ | $6.70 \times 10^{-31}$ | $1.26 \times 10^{-31}$ | $6.50 \times 10^{-31}$ |
| 7000000 | $5.86 \times 10^{-32}$ | $6.55 \times 10^{-431}$ | $4.64 \times 10^{-31}$ | $5.59 \times 10^{-31}$ |
| 8000000 | $3.40 \times 10^{-32}$ | $1.17 \times 10^{-32}$ | $8.39 \times 10^{-31}$ | $1.00 \times 10^{-32}$ |
| 9000000 | $2.03 \times 10^{-32}$ | $6.21 \times 10^{-32}$ | $1.49 \times 10^{-31}$ | $1.27 \times 10^{-32}$ |
| 10000000 | $1.46 \times 10^{-32}$ | $3.55 \times 10^{-32}$ | $9.93 \times 10^{-32}$ | $1.68 \times 10^{-32}$ |

TABLE 8. $CRB_\theta$ VERSUS PROPAGATION VELOCITY FOR AN ARRAY WITH FOUR ELEMENTS (M=4)

| Points (m/s) | Primary | | An Element Changing | |
|---|---|---|---|---|
| | 2 Sources | 3 Sources | 2 Sources | 3 Sources |
| 1000000 | $8.32 \times 10^{-35}$ | $8.92 \times 10^{-35}$ | $8.55 \times 10^{-36}$ | $4.13 \times 10^{-36}$ |
| 2000000 | $4.75 \times 10^{-35}$ | $3.96 \times 10^{-34}$ | $4.75 \times 10^{-36}$ | $3.95 \times 10^{-36}$ |
| 3000000 | $2.55 \times 10^{-35}$ | $9.54 \times 10^{-34}$ | $3.38 \times 10^{-36}$ | $5.86 \times 10^{-36}$ |
| 4000000 | $2.75 \times 10^{-35}$ | $1.24 \times 10^{-33}$ | $1.22 \times 10^{-35}$ | $2.71 \times 10^{-35}$ |
| 5000000 | $7.08 \times 10^{-34}$ | $7.45 \times 10^{-33}$ | $5.66 \times 10^{-35}$ | $3.64 \times 10^{-35}$ |
| 6000000 | $4.24 \times 10^{-34}$ | $1.98 \times 10^{-33}$ | $3.47 \times 10^{-35}$ | $3.38 \times 10^{-35}$ |
| 7000000 | $7.02 \times 10^{-34}$ | $2.01 \times 10^{-32}$ | $8.65 \times 10^{-34}$ | $4.97 \times 10^{-35}$ |
| 8000000 | $4.54 \times 10^{-34}$ | $3.39 \times 10^{-32}$ | $2.68 \times 10^{-34}$ | $2.62 \times 10^{-34}$ |
| 9000000 | $1.23 \times 10^{-33}$ | $1.51 \times 10^{-32}$ | $7.37 \times 10^{-32}$ | $5.27 \times 10^{-34}$ |
| 10000000 | $1.44 \times 10^{-33}$ | $5.43 \times 10^{-32}$ | $1.28 \times 10^{-32}$ | $3.33 \times 10^{-34}$ |

TABLE 9. $CRB_r$ VERSUS FREQUENCY FOR AN ARRAY WITH FOUR ELEMENTS (M=4)

| Points | Primary | | An Element Changing | |
|---|---|---|---|---|
| (Hz) | 2 Sources | 3 Sources | 2 Sources | 3 Sources |
| 1000000 | $2.48 \times 10^{-28}$ | $1.56 \times 10^{-25}$ | $4.83 \times 10^{-29}$ | $6.11 \times 10^{-26}$ |
| 2000000 | $8.59 \times 10^{-28}$ | $8.22 \times 10^{-26}$ | $2.58 \times 10^{-30}$ | $3.82 \times 10^{-26}$ |
| 3000000 | $4.08 \times 10^{-29}$ | $5.46 \times 10^{-26}$ | $8.76 \times 10^{-30}$ | $6.10 \times 10^{-26}$ |
| 4000000 | $2.41 \times 10^{-29}$ | $3.06 \times 10^{-27}$ | $2.31 \times 10^{-30}$ | $1.49 \times 10^{-27}$ |
| 5000000 | $2.21 \times 10^{-29}$ | $9.43 \times 10^{-27}$ | $2.15 \times 10^{-30}$ | $1.73 \times 10^{-27}$ |
| 6000000 | $1.42 \times 10^{-30}$ | $1.72 \times 10^{-27}$ | $7.72 \times 10^{-30}$ | $3.32 \times 10^{-27}$ |
| 7000000 | $1.03 \times 10^{-30}$ | $1.04 \times 10^{-28}$ | $3.41 \times 10^{-30}$ | $1.01 \times 10^{-28}$ |
| 8000000 | $2.57 \times 10^{-30}$ | $9.04 \times 10^{-28}$ | $1.18 \times 10^{-30}$ | $5.56 \times 10^{-28}$ |
| 9000000 | $1.98 \times 10^{-30}$ | $1.30 \times 10^{-29}$ | $1.36 \times 10^{-30}$ | $6.22 \times 10^{-28}$ |
| 10000000 | $1.99 \times 10^{-30}$ | $9.87 \times 10^{-29}$ | $2.01 \times 10^{-30}$ | $6.31 \times 10^{-28}$ |

TABLE 10. *CRB_r* VERSUS PROPAGATION VELOCITY FOR AN ARRAY WITH FOUR ELEMENTS (M=4)

| Points (m/s) | Primary | | An Element Changing | |
|---|---|---|---|---|
| | 2 Sources | 3 Sources | 2 Sources | 3 Sources |
| 1000000 | $2.21 \times 10^{-34}$ | $3.69 \times 10^{-32}$ | $9.07 \times 10^{-35}$ | $7.82 \times 10^{-33}$ |
| 2000000 | $6.78 \times 10^{-34}$ | $7.59 \times 10^{-32}$ | $5.64 \times 10^{-35}$ | $1.39 \times 10^{-32}$ |
| 3000000 | $8.11 \times 10^{-34}$ | $5.26 \times 10^{-31}$ | $1.18 \times 10^{-34}$ | $3.01 \times 10^{-32}$ |
| 4000000 | $2.07 \times 10^{-35}$ | $2.00 \times 10^{-30}$ | $1.48 \times 10^{-34}$ | $1.04 \times 10^{-32}$ |
| 5000000 | $4.56 \times 10^{-35}$ | $4.87 \times 10^{-30}$ | $1.15 \times 10^{-34}$ | $7.51 \times 10^{-31}$ |
| 6000000 | $8.57 \times 10^{-35}$ | $1.77 \times 10^{-29}$ | $3.07 \times 10^{-34}$ | $8.48 \times 10^{-31}$ |
| 7000000 | $4.69 \times 10^{-35}$ | $3.19 \times 10^{-29}$ | $2.72 \times 10^{-34}$ | $1.21 \times 10^{-30}$ |
| 8000000 | $6.62 \times 10^{-35}$ | $7.43 \times 10^{-29}$ | $1.74 \times 10^{-33}$ | $1.26 \times 10^{-30}$ |
| 9000000 | $4.17 \times 10^{-35}$ | $2.56 \times 10^{-28}$ | $5.16 \times 10^{-33}$ | $3.33 \times 10^{-30}$ |
| 10000000 | $9.69 \times 10^{-35}$ | $3.25 \times 10^{-28}$ | $9.25 \times 10^{-33}$ | $6.01 \times 10^{-30}$ |

CONCLUSION

This paper offers an algorithm to improve the position of an array element for any array designed on the basis of some certain or random rules. In the proposed method, one element moves on its original direction, i.e., keeping the vertical distance with each source, to reach the constellation with less CRB. The performance of this method has been demonstrated through simulation and a comparison of the receptive signals covariance matrix determinant against CRB have been made before and after the use of this method. The results of the new constellation via an element changing method has better performance and less CRB than primary constellation.

APPENDIX

*CRB equations details:*

In this appendix we can see more details of equations (13)-(18) factors. It should be noted that $\mathbf{Q}_t$ and $\mathbf{Q}_4$ are constant matrices [2]:

$$\mathbf{Q}_t = \left[ \mathbf{Q}^H \overline{\mathbf{Q}}^H \right]^H \tag{A.1}$$

$$\mathbf{Q}_4 = \mathbf{1}\left( \underline{\overline{J}}_4, \underline{J}_4 \right) \tag{A.2}$$

where

$$\mathbf{Q} = \mathbf{Q}_2 \mathbf{Q}_1 \tag{A.3}$$

$$\overline{\mathbf{Q}} = -j \overline{\mathbf{Q}}_2 \overline{\mathbf{Q}}_1 \tag{A.4}$$

$$\mathbf{Q}_1 = \mathbf{I} + \mathbf{1}\left( \underline{J}_1, \underline{J}_2 \right) \tag{A.5}$$

where $\mathbf{I}$ is the identify matrix of size equal to $N^2 \times N^2$ and $\mathbf{1}(\underline{J}_1, \underline{J}_2)$ defines a matrix, with the same size, with elements corresponding to the indexes defined by $\underline{J}_1$ and $\underline{J}_2$ equal to 1 and other elements are zero, and also we have:

$$\overline{\mathbf{Q}}_1 = \mathbf{I} - \mathbf{1}(\underline{J}_1, \underline{J}_2) \tag{A.6}$$

$$\mathbf{Q}_2 = \mathbf{1}(\overline{J}_3, \underline{J}_3) \tag{A.7}$$

$$\overline{\mathbf{Q}}_2 = \mathbf{1}(\overline{J}_1, \underline{J}_1) \tag{A.8}$$

$$\underline{J}_1 = [2, 3, \dots, N, N+3, N+4, \dots, 2N, 2N+4, 2N+5, \dots, 3N, 3N+1, \dots, (N+1)N] \tag{A.9}$$

$$\overline{J}_1 = [1, 2, \dots, (N-1)N/2] \tag{A.10}$$

$$\underline{J}_2 = [N+1, 2N+1, \dots, (N-1)N+1, 2N+2, 3N+2, \dots, (N-1)N+2, (N-1)N+3, \dots, N^2-1] \tag{A.11}$$

$$\underline{J}_3 = [1, 2, \dots, N, N+2, N+3, \dots, 2N, 2N+1, 2N+2, \dots, N^2] \tag{A.12}$$

$$\overline{J}_3 = [1, 2, \dots, N(N+1)/2] \tag{A.13}$$

$$\underline{J}_4 = [1, N+2, 2N+3, \dots, N^2] \tag{A.14}$$

and $\overline{J}_4$ is simply the indexes $[1, 2, \dots, N]$, and also

$$\dot{\mathbf{A}}_{\underline{\alpha}} = \sum_{n=1}^{N} \frac{\partial \mathbf{A}}{\partial \underline{\alpha}_n} \tag{A.15}$$

the needed derivations in the above formula are as follows:

$$\frac{\partial \mathbf{A}_{mk}}{\partial \underline{\theta}_n} = -\frac{\delta_{kn} j(\omega_0/c) \mathbf{A}_{mk} \rho_m \underline{r}_n \sin(\underline{\theta}_n - \phi_m)}{[\underline{r}^n + \rho_m^2 - 2\rho_m \underline{r}_n \cos(\underline{\theta}_n - \phi_m)]^{1/2}} \tag{A.16}$$

$$\frac{\partial \mathbf{A}_{mk}}{\partial \underline{r}_n} = -\frac{\partial \mathbf{A}_{mk}}{\partial \underline{\theta}_n} \frac{(\underline{r}_n - \rho_m \cos(\underline{\theta}_n - \phi_m))}{(\rho_m \underline{r}_n \sin(\underline{\theta}_n - \phi_m))} . \tag{A.17}$$